\documentclass[10pt,journal,compsoc]{IEEEtran}
\ifCLASSOPTIONcompsoc
  \usepackage[nocompress]{cite}
\else
  \usepackage{cite}
\fi
\ifCLASSINFOpdf
\else
\fi
\usepackage{cite}
\usepackage{amsmath,amssymb,amsfonts}
\usepackage{algorithmic}
\usepackage{graphicx}
\usepackage{subfigure}
\usepackage{textcomp}
\usepackage{xcolor}
\usepackage{booktabs}
\usepackage{footnote}
\usepackage{threeparttable}

\begin{document}
%
\title{DeepInsight: Interpretability Assisting Detection of Adversarial Samples on Graphs}
%
%
%
%

\author{Junhao~Zhu,~Yalu~Shan,~Jinhuan~Wang,~and~Shanqing~Yu,\\Guanrong~Chen,~\IEEEmembership{Fellow,~IEEE},~Qi~Xuan,~\IEEEmembership{Member,~IEEE}
\IEEEcompsocitemizethanks{
\IEEEcompsocthanksitem J. Zhu, Y. Shan, J. Wang, S. Yu, and Q. Xuan are with the Institute of Cyberspace Security, 
College of Information Engineering, Zhejiang University of Technology, Hangzhou, China (E-mail: junhao-zhu@foxmail.com; yalushan@foxmail.com; jinhuanwang@zjut.edu.cn; yushanqing@zjut.edu.cn; xuanqi@zjut.edu.cn).\protect 
\IEEEcompsocthanksitem G. Chen is with the Department of Electrical Engineering, City University of Hong Kong, Hong Kong SAR, China. (E-mail: eegchen@cityu.edu.hk).\protect
\IEEEcompsocthanksitem J. Zhu and Y. Shan have equal contributions. \protect
\IEEEcompsocthanksitem Corresponding author: Qi Xuan.}
}

%
%

\markboth{}
{Shell \MakeLowercase{\textit{et al.}}: Bare Demo of IEEEtran.cls for Computer Society Journals}
%



\IEEEtitleabstractindextext{%
\begin{abstract}
With the rapid development of artificial intelligence, a number of machine learning algorithms, such as graph neural networks have been proposed to facilitate network analysis or graph data mining. Although effective, recent studies show that these advanced methods may suffer from adversarial attacks, i.e., they may lose effectiveness when only a small fraction of links are unexpectedly changed. This paper investigates three well-known adversarial attack methods, i.e., Nettack, Meta Attack, and GradArgmax. It is found that different attack methods have their specific attack preferences on changing the target network structures. Such \emph{attack patterns} are further verified by experimental results on some real-world networks, revealing that generally the top four most important network attributes on detecting adversarial samples suffice to explain the preference of an attack method. Based on these findings, the network attributes are utilized to design machine learning models for adversarial sample detection and attack method recognition with outstanding performance.
\end{abstract}

\begin{IEEEkeywords}
Adversarial attack, adversarial defense, network structure, node classification, social network, graph data mining.
\end{IEEEkeywords}}

\maketitle

\IEEEdisplaynontitleabstractindextext

%
\IEEEpeerreviewmaketitle

\IEEEraisesectionheading{\section{Introduction}\label{sec:introduction}}

%
%
%
%
\IEEEPARstart{T}{he} vulnerability of deep learning models suffers from various adversarial attacks, which has alerted the research community of network science in recent years. Using different adversarial attack methods, adversaries can downgrade the performance of graph data mining algorithms by adding or removing some nodes or links. For example, adversaries can register fake accounts and add some fake records to mislead a recommendation system to recommend illegal products to ordinary users, leading to serious consequences~\cite{lin2020attacking,chen2020data}. 

To date, a number of studies about adversarial attacks on Graph Neural Networks (GNNs) for node classification have been reported. For example, Z$\mathrm{\ddot u}$gner et al.~\cite{zugner2019adversarial} proposed the Meta Attack against node classification. They treated input data as a hyperparameter and utilized the meta-grad to craft adversarial samples, which could downgrade the overall performance of the Graph Convolution Network (GCN) model on node classification. Chen et al.~\cite{chen2020mga} developed the Momentum Gradient Attack (MGA) taking advantage of the problem caused by poor local optimum. Besides these gradient-based attack methods, there are also many non-gradient-based attack methods. Z$\mathrm{\ddot u}$gner et al.~\cite{zugner2018adversarial} proposed an efficient algorithm named Nettack, which could generate adversarial samples with imperceptible perturbations against node classification.  Dai et al.~\cite{dai2018adversarial} utilized a reinforcement learning method to attack node or graph classification. Yu et al.~\cite{yunetwork} introduced a Genetic Algorithm (GA) based Euclidean Distance Attack (EDA) on graph embedding, successfully disturbing node embedding vectors, so as to destroy some downstream network algorithms, such as community detection and node classification.

Despite the fact that the aforementioned adversarial attack methods could significantly downgrade the performance of GNN models, it is unclear how they work and which topological attributes they really impact on. Such lack of interpretability halts the further development of adversarial attacks and the design of defense methods. Regarding this important issue, there are few studies on the analysis of adversarial attacks. Z$\mathrm{\ddot u}$gner et al.~\cite{zugner2019adversarial} compared the links inserted by Meta Attack to the links originally present in a clean network. They investigated three metrics, i.e., shortest path lengths, edge betweenness centrality, and node degree, and found that these metrics did not have significantly changes under attacks. Wu et al.~\cite{wu2019adversarial} found that many attacks tend to connect the target node to those nodes with different features and labels, and verified this observation by measuring the similarity of node features. 

\begin{figure}[!t]
    \centering
    \includegraphics[width=\linewidth]{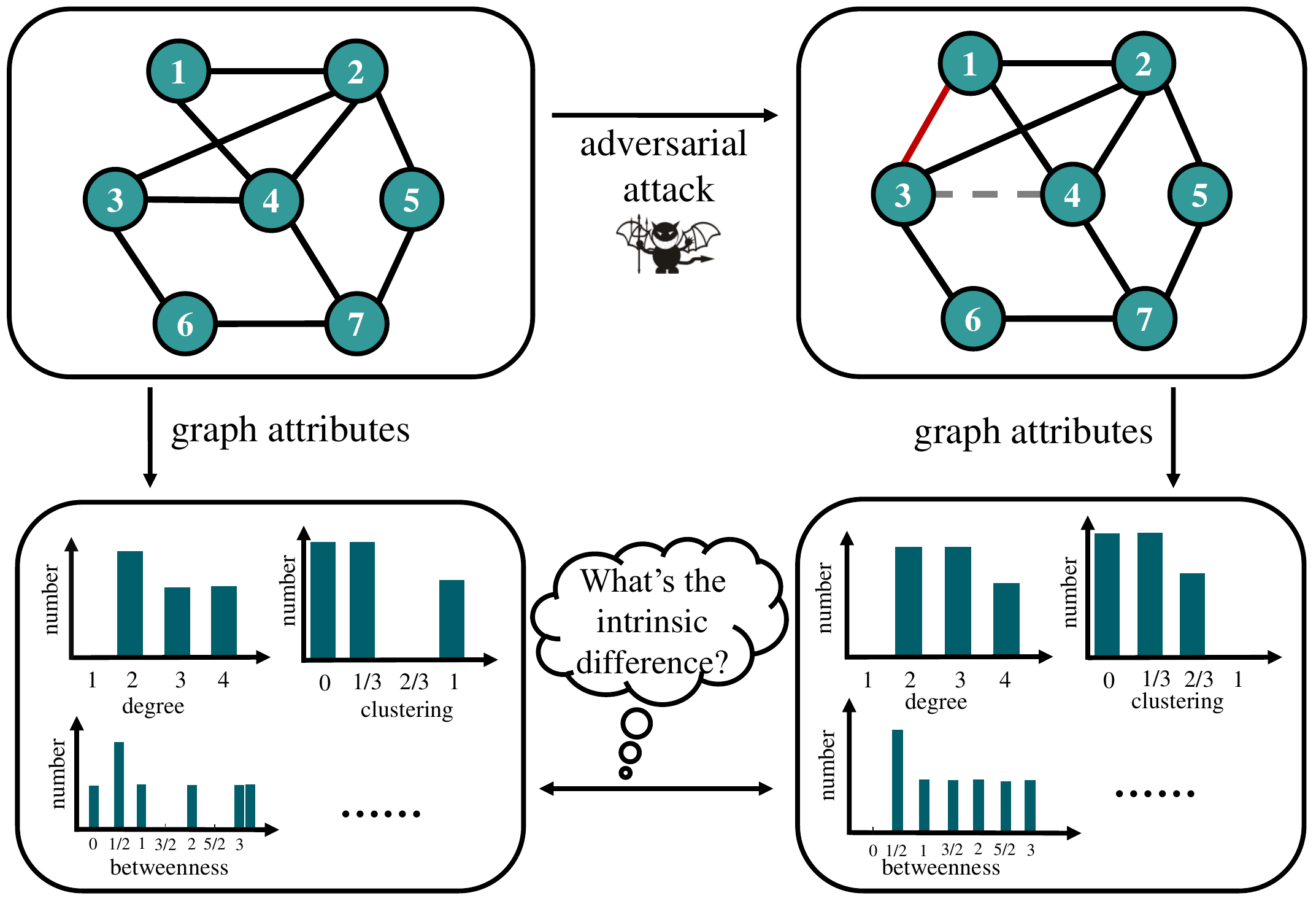}
    \caption{Adversary rewires one link to misclassify the target node, which causes a huge change in the network attributes.}
    \label{fig:overview}
\end{figure}

In order to conduct a more in-depth analysis and better explain various adversarial attack methods, this paper first infers \emph{attack patterns} by unrolling the attack procedure, to reveal different attacks for their individual sensitive attributes. In other words, a certain attack may lead to significant changes of a few network properties by rewiring a very small fraction of links. Based on these analytic results, it is possible to design effective methods to detect adversarial samples, and recognize the attack method based only on a small number of structural attributes. In Fig.~\ref{fig:overview}, a toy example is used to illustrate an adversarial attack, where one can see that although the adversary only rewires one link, a number of network attributes have changed significantly. 

The main contributions of this work are summarized as follows.
\begin{itemize}
\item \emph{Attack Pattern Inference.} To the best of our knowledge, this is the first work to associate various adversarial attacks to the variation of particular groups of network attributes, providing analytic interpretability for adversarial attacks on graphs.

\item \emph{Attack Pattern Validation.} We use 17 node-level and subgraph-level attributes to establish a Random Forest model for adversarial sample detection. We rank the attributes based on the Gini Importance, and find that the top 4 most sensitive attributes can well explain theoretically the inferred attack patterns. 

\item \emph{Adversarial Sample Detection.} 
We only select the top $k$ most sensitive attributes among all 17 attributes for adversarial detection. The method has surprisingly high accuracy, even when $k=1$, which is above 0.75, around 0.9, and even as high as 0.98, when we detect adversarial samples generated by Nettack, Meta Attack, and GradArgmax, respectively. We further use these attributes to develop machine learning models to precisely recognize attack methods, for the first time in the field, with the AUC (area under curve) larger than 0.92 when Random Forest is adopted.
\end{itemize}

The rest of paper is organized as follows. In Section~\ref{sec:relatedwork}, some related work about adversarial attacks is briefly introduced. In Section~\ref{sec:analysis}, three classical adversarial attack methods are investigated, and simple structural variation patterns of each attack method are revealed. In Section~\ref{sec:attr}, a number of network attributes are introduced, which can be ranked by Gini Importance based on the Random Forest model for adversarial sample detection. In Section~\ref{sec:experiment}, experiments are conducted, showing that the top 4 network attributes with the highest Gini Importance suffice to explain attack patterns and be used to detect adversarial samples. 
Finally, in Section~\ref{sec:conclusion}, the paper is concluded with a prospect of future work.

\section{Related Work}\label{sec:relatedwork}
A brief review  on adversarial attacks against node classification is given in the following. From the perspective of optimization, these attacks can be classified into gradient-based and non-gradient-based methods~\cite{chen2020survey}.

\subsection{Gradient-based Attack}
Overall, gradient-based attack is simple but effective. It fixes the model parameters and treats inputs as optimized hyperparameters. Using gradient-based attacks, the adversary can quickly find the links or features that have the greatest impact on model performance and then flip them to realize the attack. In relevant studies, Chen et al.~\cite{chen2018fast} considered the test-time attack (i.e. evasion attack) and proposed a Fast Gradient Attack (FGA) framework to generate adversarial graphs by utilizing the iterative gradient information of pairwise nodes based on the GCN model. Z$\mathrm{\ddot u}$gner et al.~\cite{zugner2019adversarial} adopted meta-gradients to solve the bilevel problem underlying training-time attacks (i.e. poisoning attack). The adversarial graph crafted by the Meta Attack can downgrade the overall performance of the GCN model. Wu et al.~\cite{wu2019adversarial} realized that the discrete nature of the graph data limits the adoption of existing gradient-based adversarial attacks used with non-graph data on the GCN model. To solve this problem, they proposed an integrated gradients-guided attack method, named IG-JSMA, which achieves promising attack performance. Moreover, Chen et al.~\cite{chen2020mga} developed the Momentum Gradient Attack (MGA) strategy since the momentum gradient not only can stabilize the updating process but also helps the model jump out of poor local optimum. To mention one more, Li et al.~\cite{li2020adversarial} proposed a Simplified Gradient-based Attack (SGA) to effectively attack large-scale graphs. 

\subsection{Non-gradient-based Attack}
In addition to gradient information, adversaries could generate adversarial samples in many other ways. For example, Z$\mathrm{\ddot u}$gner et al.~\cite{zugner2018adversarial} proposed an efficient algorithm named Nettack to generate adversarial samples with imperceptible perturbation. They considered both test-time and training-time attacks and used a surrogate model to tackle the bilevel optimization problem. Dai et al.~\cite{dai2018adversarial} and Ma et al.~\cite{ma2019attacking} both used the reinforcement learning-based method to generate adversarial samples. They adopted the Q-learning function and the finite horizon Markov decision process to guide the adversary to fool the target model. Chen et al.~\cite{chen2020adaptive} adjusted the traditional GAN and considered an adaptive graph adversarial attack (AGA-GAN), which consists of a multi-strategy generator, a similarity discriminator, and an attack discriminator. To reduce the cost of attack, they generated the adversarial subgraphs according to different attack strategies to rewire the corresponding parts in the original graph, and finally form the whole adversarial graph. Bose et al.~\cite{bose2019generalizable} viewed the adversarial attack problem as a generative modeling problem, and designed an encoder-decoder framework to generate adversarial samples based on the original network. 

Although the studies of adversarial attacks on graphs have blossomed in recent years, there still lacks of explanation on how these attacks work and what the intrinsic difference is between them. To address this concerned issue, the present paper investigates several well-known adversarial attacks on graphs, and associate them with the distinct changes of particular network properties under attack. Such \emph{deep insights} can facilitate the detection of adversarial samples and improve the recognition of adversarial attacks based only on several network attributes. 

\section{Attack Pattern Inference}\label{sec:analysis}
In recent years, there has been extensive research on adversarial attacks against node classification. Among them, Nettack, Meta Attack, and GradArgmax have received particular attention, and are often used as baselines. Therefore, this paper focus on these three typical adversarial attack methods, and investigates their attack goals and strategies to discover the underlying principles.

\subsection{Nettack}\label{subsec:Nettack}
Adversaries often regard the GCN model as their attack target, due to its outstanding performance on node classification. Traditionally, a GCN model with two layers is represented by:

\begin{equation}
    \mathbf{z} = softmax(\hat{A} \sigma (\hat{A}XW^{(1)})W^{(2)}),
    \label{gcn}
\end{equation}
where $\hat{A}=D^{-\frac{1}{2}}\widetilde{A}D^{-\frac{1}{2}}$, $\widetilde{A}=I+A$ is the adjacency matrix of graph $G$ after adding self-loops via the identity matrix $I$, $D$ is a diagonal matrix with $D_{ii} = \sum_{j}{\widetilde{A}_{ij}}$, $W^{(l)}$ is the trainable weight matrix of layer $l$, $\sigma (\cdot)$ is an activation function, e.g., ReLU and $X$ is the node feature matrix.

Z$\mathrm{\ddot u}$gner et al.~\cite{zugner2018adversarial} considered the training-time attack, i.e., poisoning attack, which can be mathematically formulated as a bilevel optimization problem:
\begin{equation}
\begin{aligned}
    & \mathop{\arg\max}_{A^{'}\in \Phi(A)} \max_{y\neq y_{old}} \ln{\mathbf{z}^{*}_{\mu,y}} - \ln{\mathbf{z}^{*}_{\mu,y_{old}}} \\
    s.t.~ & \mathbf{z}^{*}=f_{\theta^*}(A^{'},X) \mathrm{~with~} \theta^*=\mathop{\arg\max}_{\theta}\mathcal{L}(\mathbf{z}, y_{old}; \theta),
\end{aligned} \label{nettack}
\end{equation}
where $\Phi(A)$ represents the set of adversarial adjacency matrix, and $y_{old}$ and $y$ denote the class for the target node $\mu$ based on the original graph and the adversarial one, respectively, and $\mathcal{L}(*)$ is the loss function that is instantiated with a cross entropy loss. 

To improve transferability, they performed the attack on a surrogate model, a two-layer GCN model. In order to simplify the surrogate model, they instantiated $\sigma (\cdot)$ with a simple linear activation function, i.e., $ \mathbf{y} = softmax(\hat{A}^{2}XW) $. 

Since the attack goal is to maximize the difference in the log-probabilities of the target node as shown in Eq.~(\ref{nettack}), the instance-dependent normalization induced by the softmax can be ignored. Thus, the log-probability of the target node can be denoted as $\hat{A}^{2}XW$, with the goal re-written as
\begin{equation}
\begin{aligned}
    & \max_{y\neq y_{old}} \ln{\mathbf{z}^{*}_{\mu,y}} - \ln{\mathbf{z}^{*}_{\mu,y_{old}}} \\
  \Leftrightarrow & \max [\hat{A^{'}}^2XW]_{\mu,y}-[\hat{A^{'}}^2XW]_{\mu,y_{old}}\\
  :\Leftrightarrow &\max [\hat{A^{'}}^2C]_{\mu,y}-[\hat{A^{'}}^2C]_{\mu,y_{old}}.
\end{aligned}\label{ntkDef}
\end{equation}
Now, substitute $XW$ with $C := XW \in\mathbb{R}^{N\times K}$, since the parameters of the surrogate model are fixed, which means that $XW$ is constant.
For ease of understanding, $\hat{A}$ is unrolled to another notation: $\hat{A}=\left( \frac{a_{ij}}{\sqrt{d_{i}d_{j}}} \right)_{N\times N}$, where $d_i$ is the degree of node $i$ and $a_{ij}$ is the entry at position $(i,j)$ in the matrix $\hat{A}$. One can derive
\begin{equation*}
    \hat{A}^2=\left( \frac{1}{\sqrt{d_{i}d_{j}}}\sum_{k=1}^N \frac{a_{ik}a_{jk}}{d_k}\right)_{N\times N}.
\end{equation*}

The log-probability of the target node $\mu$ obtained from the surrogate model is then represented as 
\begin{equation}
\begin{aligned}
   \hat{A}^2C|_{\mu,y}=  \left( \frac{1}{\sqrt{d_{\mu}d_{j}}}\sum_{k=1}^N \frac{a_{\mu k}a_{j k}}{d_k}
\right)_{1\times N} \cdot
  \begin{bmatrix}
    c_{1y} \\ c_{2y} \\ \vdots \\c_{Ny}
  \end{bmatrix} \\
  = \sum_{j=1}^N \left(\frac{c_{jy}}{\sqrt{d_{\mu}d_{j}}}\sum_{k\in \mathcal{CN}(\mu,j)\cup\{\mu\}\cup\{j\}} \frac{1}{d_k} \right),
\end{aligned}\label{hat_A2C}
\end{equation}
where $c_{iy}$ is an entry of $C$, and $\mathcal{CN}(\mu,v)$ is the set of the common neighbors of node $\mu$ and node $v$. If $\mu=v$, then $\mathcal{CN}(\mu,v)$ is equivalent to the set of neighbors of node $\mu$.

Similarly, $[\hat{A}^{2}C]_{\mu,y_{old}}$ can be represented as Eq.~(\ref{hat_A2C}). So, the attack goal of the Nettack could be further re-written as
\begin{equation}
\begin{aligned}
  &\max [\hat{A}^2C]_{\mu,y} - [\hat{A}^2C]_{\mu,y_{old}} \\ \Leftrightarrow &\max\left\{ \sum_{j=1}^N \frac{c_{jy}-c_{jy_{old}}}{\sqrt{d_{\mu}d_{j}}} \sum_{k\in \mathcal{CN}(\mu,j)\cup\{\mu\}\cup\{j\}} \frac{1}{d_k} \right\}.
\end{aligned}\label{ntk_cal}
\end{equation}

\noindent \textbf{Deep Insight.} According to Eq.~(\ref{ntk_cal}), one can get two simple attack patterns: 1) modify links between nodes with small degrees; and 2) increase the number of common neighbors of the target node and other nodes. Furthermore, one can see that Nettack is more likely to add new small-degree neighbors to the target node. Considering that two attack patterns are both about node degree and common neighbors, it is reasonable to speculate that Nettack is apt to modify degree-based or common-neighbor-based topological attributes, e.g., node degree and clustering coefficient.

\subsection{Meta Attack}\label{subsec:meta}
Inspired by meta-learning \cite{bengio2000gradient}, Z{\"u}gner et al. \cite{zugner2019adversarial} proposed a meta-gradient-based approach to attak the GCN model for node classification, named Meta Attack. Different from Nettack, Meta Attack tackles a bilevel optimization problem by turning the gradient-based optimization procedure of node classification upside down and treating the input graph data as a hyperparameter to learn. The goal of Meta Attack is defined as
\begin{equation}
\begin{aligned}
 & \min_{\hat{G}\in \Phi(G)} \mathcal{L}_{atk}(f_{\theta^{*}}(\hat{G})) \\
 s.t. & \theta^{*}=\mathop{\arg\min}_{\theta}\mathcal{L}_{train}(f_{\theta}(\hat{G})),
\end{aligned} \label{mtkDef}
\end{equation}
where $\Phi(G)$ is a set of adversarial graphs.
According to the attack loss, two meta attack methods are proposed. One treats the modified training loss as the attacker loss, while the other uses the test set and the predicted labels to calculate the attacker loss. Here, the focus is on the latter, that is, $\mathcal{L}_{atk}=-\mathcal{L}_{test}$.

Thus, adversaries get the meta-gradient of each link in the graph $G$, expressed as
\begin{equation}
\begin{aligned}
\nabla_{G}^{meta}&=\nabla_{G}\mathcal{L}_{atk}(f_{\theta_t}(G))\\
& = \nabla_f\mathcal{L}_{atk}(f_{\theta_t}(G))\cdot [\nabla_Gf_{\theta_t}(G)+\nabla_{\theta_t}f_{\theta_t}(G)\cdot \nabla_G\theta_t],
\end{aligned}\label{meta-grad}
\end{equation}
with
\begin{equation}
\nabla_G \theta_{t+1} = \nabla_G\theta_t - \alpha\nabla_G\nabla_{\theta_t}\mathcal{L}_{train}(f_{\theta_t}(G)).
\label{meta-grad1}
\end{equation}
The adversaries greedily pick the perturbation $(\mu,v)$ with the highest meta-gradient at every step, i.e., 
\begin{equation}
    (\mu,v) = \mathop{\arg\max}_{(\mu, v)}\nabla^{meta}_{a_{\mu v}}\cdot(-2\cdot a_{\mu v}+1),
\end{equation} 
and then insert the link $e = (\mu, v)$ by setting $a_{\mu v} = 1$ if nodes $(\mu, v)$ are currently not connected, while delete the link by setting $a_{\mu v} = 0$ otherwise.

According to the above analysis, the adversaries choose the modified link based on $\nabla^{meta}_{a_{\mu v}}$. Let $\mathbf{y}$ denote the surrogate model prediction result, and $\mathbf{y^{*}}$ be the true label of the node.
Given that $\nabla_{G}\theta_t$ is determined by training loss, it is a constant when the gradients of loss are unrolled on the test set. So, $\nabla_{G}^{meta}$ can be simplified as

\begin{equation}
\begin{aligned}
\nabla^{meta}_{a_{\mu v}} &\propto \nabla_f\mathcal{L}_{atk}(f_{\theta_t}(G))\cdot \nabla_{G}f_{\theta_t}(G) \\
&\propto \frac{\partial \mathcal{L}(\mathbf{y},\mathbf{y^{*}})}{\partial \mathbf{y}}\cdot\frac{\partial \mathbf{y}}{\partial \hat{A}^2C}\cdot \frac{\partial \hat{A}^2C}{\partial a_{\mu v}},
\end{aligned}\label{mtk_cal}
\end{equation}
where $\frac{\partial \mathcal{L}(\mathbf{y},\mathbf{y^{*}})}{\partial \mathbf{y}}\cdot\frac{\partial \mathbf{y}}{\partial \hat{A}^2C}$ actually is the derivative of a softmax function with cross-entropy loss. It is known that the derivative of the softmax function with cross-entropy loss is equivalent to the difference between predicted value and its label. As a result, one has
\begin{equation}
  \nabla^{meta}_{a_{\mu v}} \propto (\mathbf{y} - \mathbf{y^{*}}) \cdot \frac{\partial \hat{A}^2C}{\partial a_{\mu v}} \label{meta}.
\end{equation}

\noindent \textbf{Deep Insight.} According to Eq.~(\ref{meta}), $\nabla^{meta}_{a_{\mu v}}$ is proportional to $\nabla_{a_{\mu v}}\hat{A}^2C$. That is, Meta Attack and Nettack have similar attack patterns, both of which realize their attack goals by adding links between the target node and the nodes with small degrees and increasing the number of their common neighbors. The difference is that Meta Attack adopts gradient information (fastest direction of reaching target) to search the best attack strategy step by step, while Nettack does not. Therefore, Meta Attack may have a greater impact on the same group of graph attributes.

\subsection{GradArgmax}
Many adversarial attacks are gradient-based. Adversaries try to get or estimate the gradient information to find the most important link to the model. In this paper, the GradArgmax attack method is adopted against a two-layer GCN model, with its goal defined as
\begin{equation}
max \quad \mathcal{L}(\mathbf{z}_{\mu}, \mathbf{y^{*}}),
\end{equation}
where $\mathbf{z}_{\mu}$ is a node embedding for the target node generated by a two-layer GCN model. Usually, the adversaries adopt cross-entropy loss, so one can calculate the partial derivative of $\mathcal{L}(\cdot)$ with respect to the adjacency matrix, so as to get the gradient information:

\begin{equation}
\frac{\partial \mathcal{L}(\mathbf{z}_{\mu}, \mathbf{y^{*}})}{\partial a_{ij}}=\sum_{k=1}^K \frac{\partial \mathcal{L}(\mathbf{z}_{\mu}, \mathbf{y^{*}})}{\partial \mathbf{z}_{\mu}}\cdot \frac{\partial \mathbf{z}_{\mu}}{\partial a_{ij}}, 
\label{grad}
\end{equation}
where $\frac{\partial \mathcal{L}(\mathbf{z}_{\mu}, \mathbf{y^{*}})}{\partial \mathbf{z_{\mu}}}$ could be viewed as the derivative of the softmax with cross-entropy loss function, which is the difference between the label and predicted value and it is a constant in the attack procedure. That is, $\frac{\partial \mathcal{L}(\mathbf{z}_{\mu}, \mathbf{y^{*}})}{\partial a_{ij}}$ is proportional to $\frac{\partial \mathbf{z}_{\mu}}{\partial a_{ij}}$. 

GradArgmax adopts the greedy algorithm to select a set of node pairs with the largest absolute value of gradient to modify. At the $t$ step of attack, the link $(i_t, j_t)$ will be added if $\frac{\partial \mathcal{L}}{\partial a_{i_t j_t}} < 0$ or it will be deleted if $\frac{\partial \mathcal{L}}{\partial a_{i_t j_t}} > 0$. To ensure that the modification is imperceptible, only a small number of links are allowed to modify. Meanwhile, in Eq.~(\ref{gcn}), the nonlinear activation $\sigma(\cdot)$ could be replaced by the ReLU function. It is known that gradients of the nonlinear activation functions ReLU will approach or equal to 1 if the input of activation functions is appropriate; otherwise, the gradients will disappear. Apparently, one only need to care about the non-zero part and ignore the cases where the gradients disappear. Consequently, one has

\begin{equation}
\begin{aligned}
\frac{\partial \mathbf{z}_{\mu}}{\partial a_{ij}} &= \frac{\partial \hat{A}}{\partial a_{ij}} \sigma(\hat{A}XW^{(1)})W^{(2)}+\hat{A}W^{(2)}\frac{\partial \sigma(\hat{A}XW^{(1)})}{\partial a_{ij}} \\
& = \frac{\partial \hat{A}}{\partial a_{ij}} \sigma(\hat{A}XW^{(1)})W^{(2)} + \hat{A}W^{(2)}XW^{(1)}\frac{\partial \hat{A}}{\partial a_{ij}} \\
& \propto \frac{\partial \hat{A}}{\partial a_{ij}} \propto \left(\frac{1}{\sqrt{d_{i}d_{j}}}\right)_{N\times N}.
\end{aligned}\label{gradAgm}
\end{equation}

\noindent \textbf{Deep Insight.} Eq.~(\ref{gradAgm}) implies that the smaller the node degree is, the larger the gradient is. That is, GradArgmax tends to attack those nodes with small degrees. In addition, Eq.~(\ref{gradAgm}) also shows that adversaries can realize the attack by modifying the links between the nodes of small degree. So GradArgmax may significantly affect degree-based attributes, such as the node degree, and further the degree of neighbors, etc. 

\section{Attack Pattern Validation} \label{sec:attr}
In Section~\ref{sec:analysis}, the three well-known adversarial attack methods on graphs were reviewed and discussed, inferring the possible sensitive structural attributes under attacks. Here, a number of node-level and subgraph-level network attributes are used to establish a classification model for adversarial sample detection. In particular, the \emph{Random Forest} classifier is adopted, with the \emph{Gini Importance} to measure the importance of each attribute in the detection. These empirical results could help verify the theoretical inference, and further, reveal the most sensitive graph attributes to a particular adversarial attack method.

\subsection{Graph Attributes}
In order to generalize of the method, one needs as many graph attributes as possible. In this work, two groups of graph attributes are extracted which complement each other: node-level attributes and subgraph-level attributes. Node-level attributes focus on the target node and the subgraph-level attributes can merge the 2-hop neighbor information. These graph attributes are widely used in node classification, link prediction, and graph classification ~\cite{xuan2019subgraph,fu2018link}.

\subsubsection{Node-level Attributes}
In a network, there are many node attributes, which are widely used in node ranking and classification. Here, six node attributes are considered, including:

\begin{itemize}
  \item \textbf{Degree ($D_i$):} Degree is the number of the links connected to a node. 
  \item \textbf{Clustering Coefficient ($C_i$):} Clustering Coefficient \cite{borgatti1997network,lind2005cycles} measures the degree to which the neighbors of a node tend to connect to each other. It is defined as 
  \begin{equation}
  C_i = \frac{2L_i}{D_i(D_i-1)}, \label{$C$}
  \end{equation}
  where $L_i$ represents the existing number of links between the neighbors of concerned node $i$, and $\frac{D_i(D_i-1)}{2}$ denotes the maximum possible number of links among its neighbors. 

  \item \textbf{Betweenness Centrality ($BC_i$):} Betweenness Centrality \cite{wasserman1994social} is a metric based on shortest paths and reveals the concerned node's ability to control the transmission of information along the shortest path between node pairs in the network. It is defined as
  \begin{equation}
  BC_i = \sum_{s\neq i\neq t}\frac{n_{st}^i}{g_{st}},\label{BC}
  \end{equation}
  where $g_{st}$ represents the number of all the shortest paths from node $s$ to node $t$, and $n_{st}^i$ denotes the number of the shortest paths from node $s$ to node $t$ which pass the node $i$.

  \item \textbf{Closeness Centrality ($CC_i$):} Closeness Centrality \cite{bavelas1950communication,beauchamp1965improved} is defined as the reciprocal of the sum of the lengths of the distances between the concerned node and all other nodes in the network. It is defined as
  \begin{equation}
  CC_i = \frac{N}{\sum_{j=1}^{N}d_{ij}},\label{CC}
  \end{equation}
  where $N$ denotes the total number of nodes in the network and $d_{ij}$ represents the distance between node $i$ and node $j$. The node with a higher closeness centrality has a greater impact on the transmission of information in the network.

  \item \textbf{Eigenvector Centrality ($EC_i$):} Eigenvector Centrality \cite{bonacich1972factoring,bonacich2007some} is a measure of the influence of a node in the graph. A node's importance depends on the number and the importance of its neighbor nodes. Eigenvector Centrality makes full use of neighbor node information and is defined as
  \begin{equation}
  EC_i = c\sum_{j=1}^N a_{ij}x_{j},\label{EC}
  \end{equation}
 where $a_{ij}$ is an entry of the adjacent matrix, and $x_j$ is an entry of a vector of node centralities $\mathbf{x}=[x_1,x_2,\cdots,x_N]^{\top}$, satisfying
  \begin{equation}
  \mathbf{x}^{(t+1)} = A\mathbf{x}^{(t)}.
  \end{equation}
  The vector of node centralities in Eq.~(\ref{EC}) is obtained after $t$ iterations. One may also start off this power iteration with initial the vector $\mathbf{x}=[1,1,\cdots,1]^{\top}$.

  \item \textbf{Average Neighbor Degree of Node ($ND_i$):} Average neighbor degree measures the average degree of all the neighbors, defined as
  \begin{equation}
  ND_i = \frac{1}{|N(i)|}\sum_{j\in N(i)}D_j, \label{ND}
  \end{equation}
  where $D_i$ is the degree of node $i$ and $N(i)$ is the set of the all neighbors of node $i$.
\end{itemize}

\subsubsection{Subgraph-level Attributes}
Inspired by the fact that the classic GCN exploits the neighbors' information, the subgraph attributes are used to represent a node. For each node, one can extract a subgraph that consists of two-hop neighbors of the target node from the original graph. Here, subgraph-level attributes are used to realize graph classification~\cite{xuan2019subgraph}. These subgraph-level attributes include:

\begin{itemize}
  \item \textbf{Number of nodes ($N_{sg}$):} The total number of nodes in the subgraph. 
  \item \textbf{Number of links ($E_{sg}$):} The total number of links in the subgraph. 
  \item \textbf{Average degree ($D_{sg}$):} The average degree of all nodes in the subgraph.
  \item \textbf{Percentage of leaf nodes ($P_{sg}$):} Nodes with only one neighbor are defined as leaf nodes. Percentage of leaf nodes is calculated by:
  \begin{equation}
    P_{sg} = \frac{F_{sg}}{N_{sg}}, \label{PLN}
  \end{equation}
  where $F_{sg}$ denotes the number of leaf nodes in the subgraph.
  
  \item \textbf{Largest eigenvalue of the adjacent matrix ($EV$):} Let $A_{sg}$ denote the adjacent matrix of the subgraph. The largest eigenvalue of $A_{sg}$ is used as one of graph-level attributes.
  
  \item \textbf{Network density ($DS$):} Let $DS$ denote the network density. Given the number of nodes $N_{sg}$ and links $E_{sg}$, the network density is defined as
  \begin{equation}
    DS = \frac{2E_{sg}}{N_{sg}(N_{sg}-1)}.\label{density}
  \end{equation}
  
  \item \textbf{Average clustering coefficient ($C_{sg}$):} Average clustering coefficient is the mean clustering coefficient of all nodes in the subgraph. It is defined as 
  \begin{equation}
    C_{sg} = \frac{1}{N_{sg}}\sum_{i\in\mathcal{N}_{sg}(\mu)}C_i, \label{aver_C}
  \end{equation}
  where $\mathcal{N}_{sg}(\mu)$ denotes the set of nodes in the subgraph, and the clustering coefficient $C_i$ is formulated in Eq.~(\ref{$C$}).
  
  \item \textbf{Average betweenness centrality ($BC_{sg}$):} Average betweenness centrality is the mean betweenness centrality of all nodes in the subgraph. It is defined as 
  \begin{equation}
    BC_{sg} = \frac{1}{N_{sg}}\sum_{i\in\mathcal{N}_{sg}(\mu)}BC_i, \label{aver_abc}
  \end{equation}
  where $BC_i$ is formulated in Eq.~(\ref{BC}).

  \item \textbf{Average closeness centrality ($CC_{sg}$):} Average closeness centrality is the mean closeness centrality of all nodes in the subgraph. It is defined as
  \begin{equation}
    CC_{sg} = \frac{1}{N_{sg}}\sum_{i\in\mathcal{N}_{sg}(\mu)}CC_i, \label{aver_acc}
  \end{equation}
   where $CC_i$ is formulated in Eq.~(\ref{CC}).
  \item \textbf{Average eigenvector centrality ($EC_{sg}$):} 
  Average eigenvector centrality is the mean closeness centrality of all nodes in the subgraph. It is defined as
  \begin{equation}
    EC_{sg} = \frac{1}{N_{sg}}\sum_{i\in\mathcal{N}_{sg}(\mu)}EC_i, \label{aver_aec}
  \end{equation}
   where $EC_i$ is formulated in Eq.~(\ref{EC}).
   
  \item \textbf{Average neighbor degree of Graph ($ND_{sg}$):} Average neighbor degree is the mean neighbor degree of all nodes in the subgraph. It is defined as
  \begin{equation}
    ND_{sg} = \frac{1}{N_{sg}}\sum_{i\in\mathcal{N}_{sg}(\mu)}ND_i, \label{aver_and}
  \end{equation}
  where $ND_i$ is formulated in Eq.~(\ref{ND}).
\end{itemize}

\subsection{Attribute Selection}
\begin{figure}[!t]
    \centering
    \includegraphics[width=3.5in]{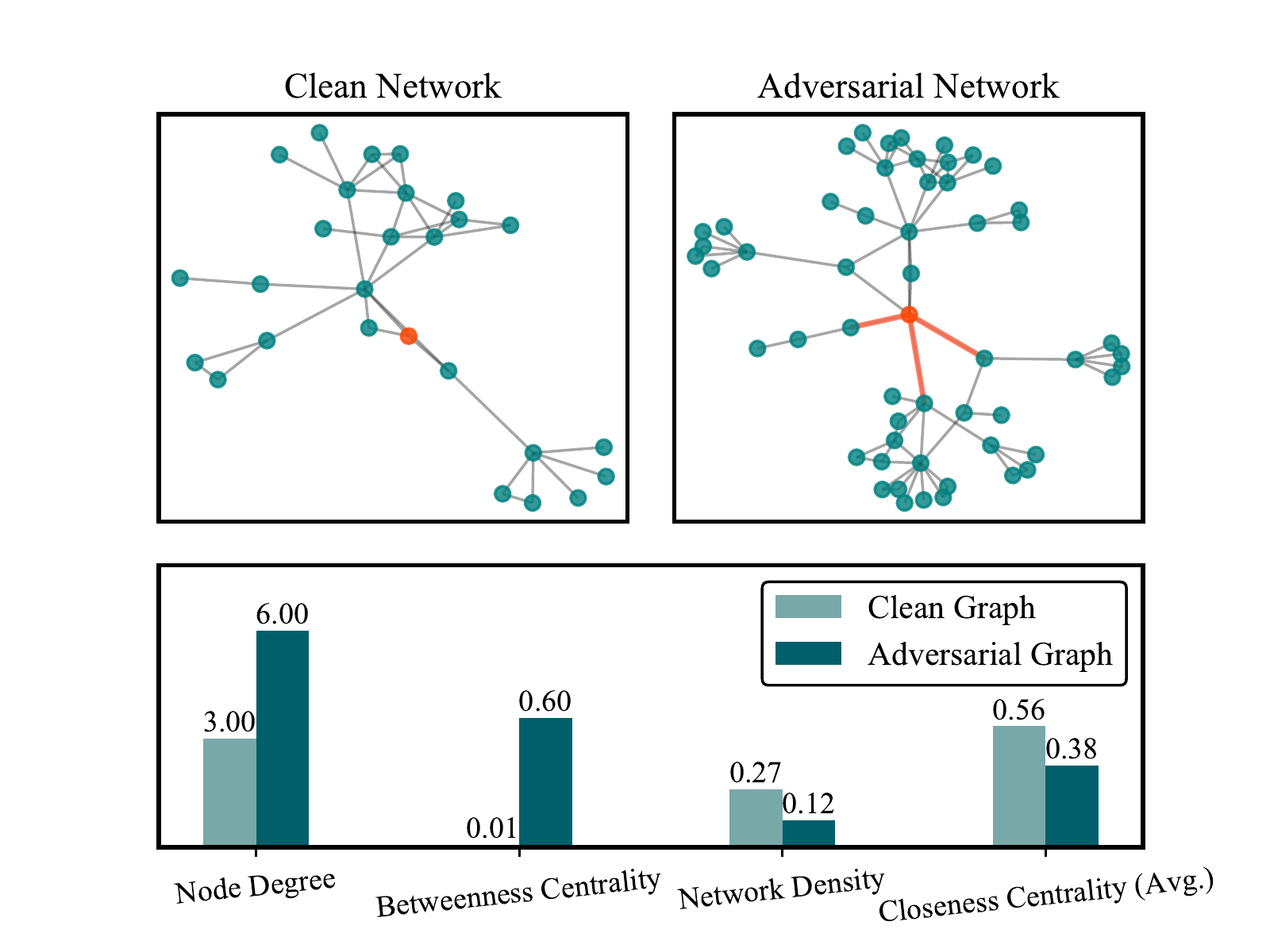}
    \caption{Comparison between a 3-hop subgraph (upper left) and its perturbed 3-hop subgraph by Nettack (upper right) in the Cora network. The concerned node and added links are marked in red. The bottom shows the difference of 4 attributes in the clean graph and adversarial graph.}
    \label{ntk_cora_example}
\end{figure}

Indeed, the graph attributes could significantly change as the network structure is modified by adversarial attack. As a simple case study, Fig.~\ref{ntk_cora_example} shows the 3-hop subgraph of the clean graph and that of the adversarial graph crafted by Nettack, and summarizes 4 attributes for comparison. The result indicates that although only 3 links have been added, the graph attributes have tremendous changes. One can reveal the attack patterns by comparing the attributes of the original graph and that of the adversarial graph. 

To accurately reveal the attack patterns, one needs to select the top $k$ attributes, representing the most sensitive to the particular attack method, from the 17 indicators summarized above. In particular, one needs to measure the importance of the attributes by utilizing \emph{Gini Importance} computed from the \emph{Random Forest} structure~\cite{1522531,breiman2001random,menze2009comparison}. Random Forest is a set of Decision Trees, and each Decision Tree is a set of internal nodes and leaves. For the internal nodes, the selected attributes are used to make a decision on how to divide samples into two categories: clean samples and adversarial samples. The attributes for internal nodes are selected with Gini impurity, which measures the likelihood that two random samples with the same attribute belong to different categories. 

High Gini impurity indicates that the attributes of adversarial samples and clean samples are similar. In other words, high Gini impurity also means that the attribute is not sensitive to adversarial attacks. One can measure how each attribute decreases the impurity of the split and the attribute with the highest decrease is selected as the internal node. For each attribute, one can collect information of how on average it decreases the impurity. The average trees in the forest are the measure of the attribute importance, i.e., Gini Importance \cite{breiman2001random}. Thereby, attributes can be ranked according to Gini Importance, and the top $k$ attributes, which contain the key information of attack patterns, can be selected to perform the subsequent analysis.

\section{Experiments and Results} \label{sec:experiment}

In this section, experiments are performed for answering the following questions:
\begin{itemize}
    \item [\textbf{RQ1}] Can the top $k$ network attributes with the highest Gini Importance based on the Random Forest model explain the attack patterns inferred in Section~\ref{sec:analysis}?
    \item [\textbf{RQ2}] Do the top $k$ attributes contain enough information to distinguish adversarial samples from normal ones? In other words, without using all the 17 attributes, can one get similar detection performance by using only the top $k$ attributes?
    \item [\textbf{RQ3}] Can one use these attributes to detect adversarial samples, and further recognize the attack method, with reasonable accuracy?
\end{itemize}

Next, the experimental settings are described, and then the above questions will be answered.

\begin{table}[htbp]
\caption{The basic properties of the four networks.}
\begin{center}
\begin{tabular}{c|c|c|c}
\toprule[1pt]
\textbf{Datasets} & \textbf{\# Nodes} & \textbf{\# Links} & \textbf{\# Classes} \\
\hline
Citeseer & 3327 & 4732 & 6 \\
\hline
Cora & 2709 & 5429 & 7 \\
\hline
Pubmed & 19717 & 44338 & 3 \\
\hline
Polblogs & 1490 & 19025 & 2 \\
\bottomrule[1pt]
\end{tabular}
\label{dataset}
\end{center}
\end{table}

\begin{figure*}[!th]
\centering
\includegraphics[width=7in]{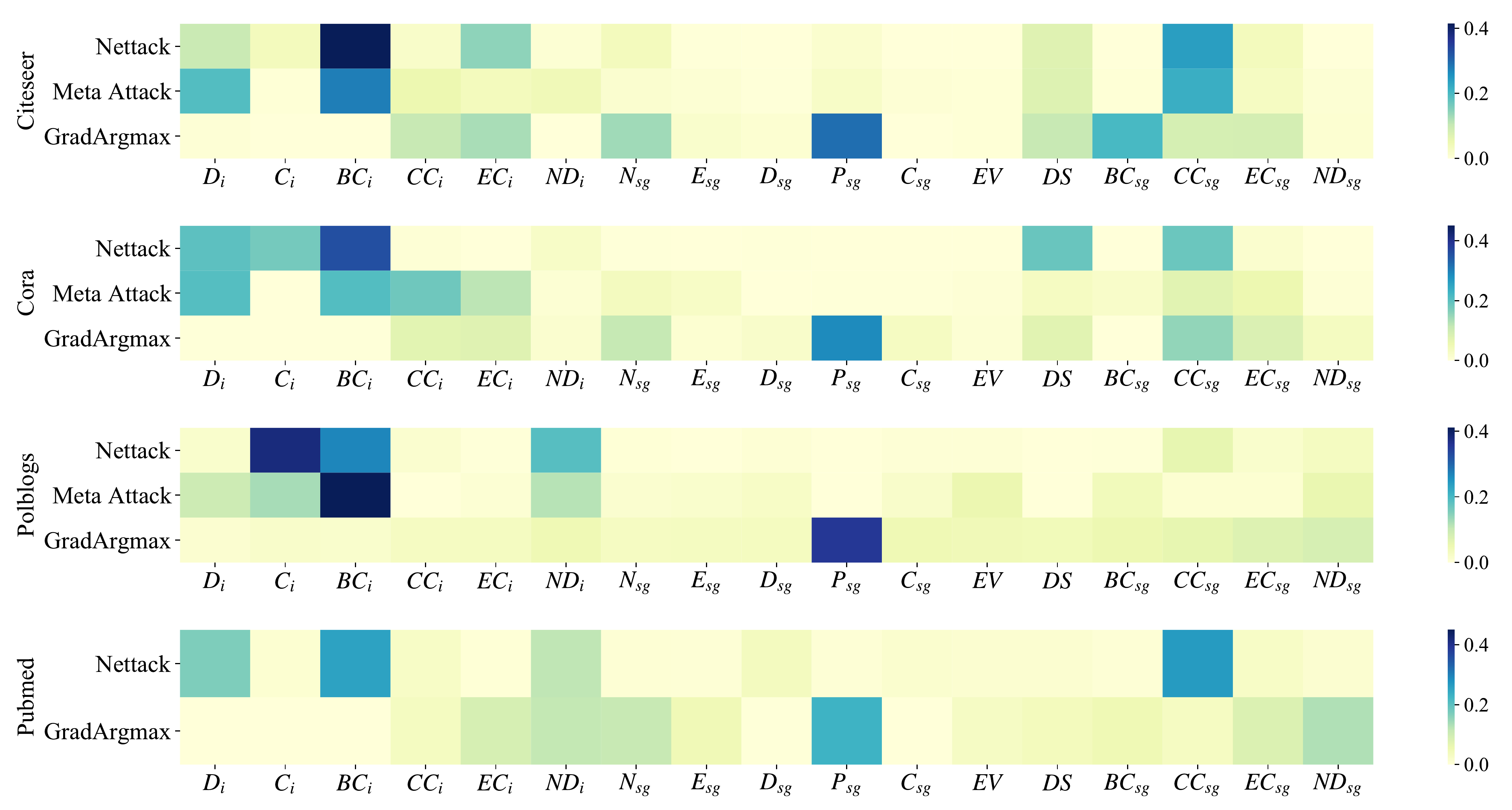}
\caption{The Gini Importance of graph attributes in Random Forest classifier. Nettack, Meta Attack, and GradArgmax are applied on the Citeseer, Cora, Polblogs, and Pubmed, respectively.}
\label{gini_impr}
\end{figure*}

\begin{table*}[!ht]
\caption{The top $4$ most sensitive attributes with the largest Gini Importance for each attack method on each network.}
\label{top3select}
\centering
\begin{threeparttable}
  \begin{tabular}{c|c|c|c|c}
    \toprule[2pt]
    Attack Methods & Citeseer & Cora & Polblogs & Pubmed \\
    \hline
    Nettack & $BC_i$ , $CC_{sg}$, $EC_i$, $D_{i}$ & $BC_i$ , $D_i$, $DS$, $CC_{sg}$ & $C_i$, $BC_i$, $ND_i$, $CC_{sg}$ & $BC_i$, $CC_{sg}$, $D_i$, $ND_{i}$\\
    \hline
    Meta Attack & $BC_i$, $CC_{sg}$, $D_i$, $DS$ & $BC_i$ , $D_i$ , $CC_i$, $EC_i$ & $BC_i$, $C_{i}$, $ND_{i}$, $D_{i}$ & - \\
    \hline
    GradArgmax & $P_{sg}$, $BC_{sg}$, $N_{sg}$, $EC_i$ & $P_{sg}$, $CC_{sg}$, $N_{sg}$, $EC_i$ & $P_{sg}$, $ND_{sg}$, $EC_{sg}$, $CC_{sg}$ & $P_{sg}$, $ND_{sg}$, $ND_i$, $N_{sg}$\\
    \bottomrule[2pt]
  \end{tabular}
  \begin{tablenotes}
    \item ``-" in this table means that computing capability needed by the adversaries exceeds the existing level of the lab device.
  \end{tablenotes}
 \end{threeparttable}
\end{table*}

\subsection{Experimental Settings}
We conducted experiments on four real-world networks: Citeseer~\cite{sen2008collective}, Cora~\cite{sen2008collective}, Pubmed~\cite{sen2008collective}, and Polblogs~\cite{adamic2005political}. The first three are citation networks where nodes correspond to papers and links indicate the citing relationship among them. The last one is a social network where nodes correspond to the blogs and links represent interactive relationships among blogs. The specifications of these datasets are given in TABLE \ref{dataset}.

\begin{figure*}[!t]
\centering
\subfigure[Nettack]{
\begin{minipage}[t]{\linewidth}
\centering
\includegraphics[width=7in]{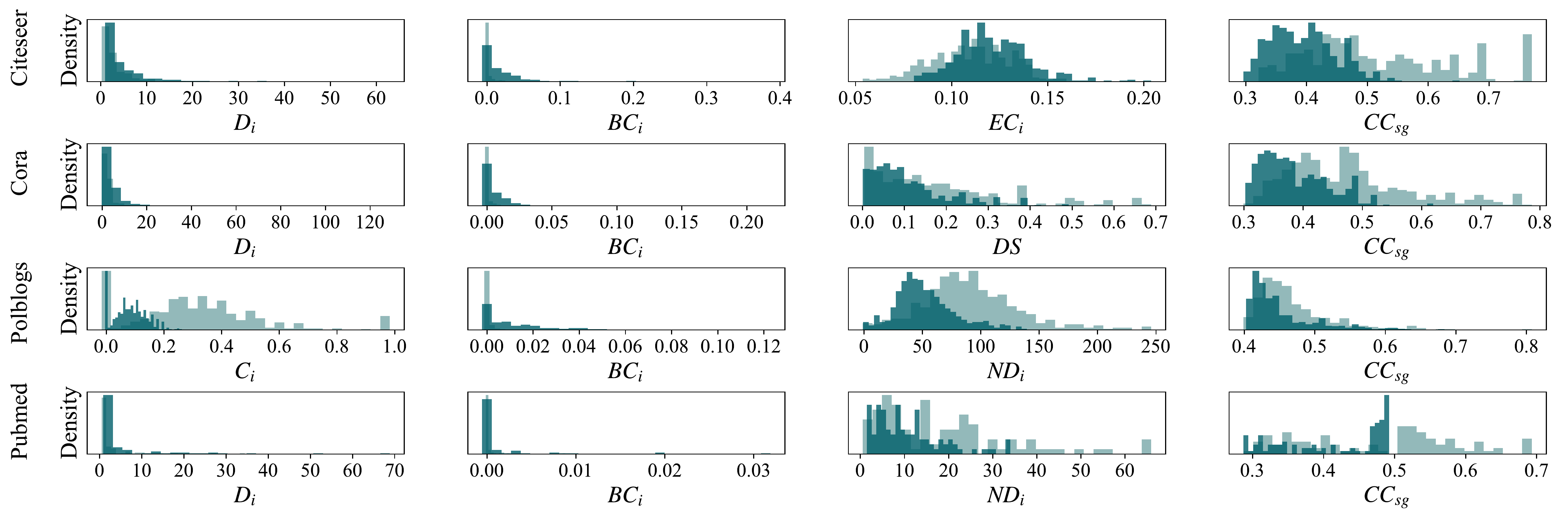}
\end{minipage}%
}%

\subfigure[Meta Attack]{
\begin{minipage}[t]{\linewidth}
\centering
\includegraphics[width=7in]{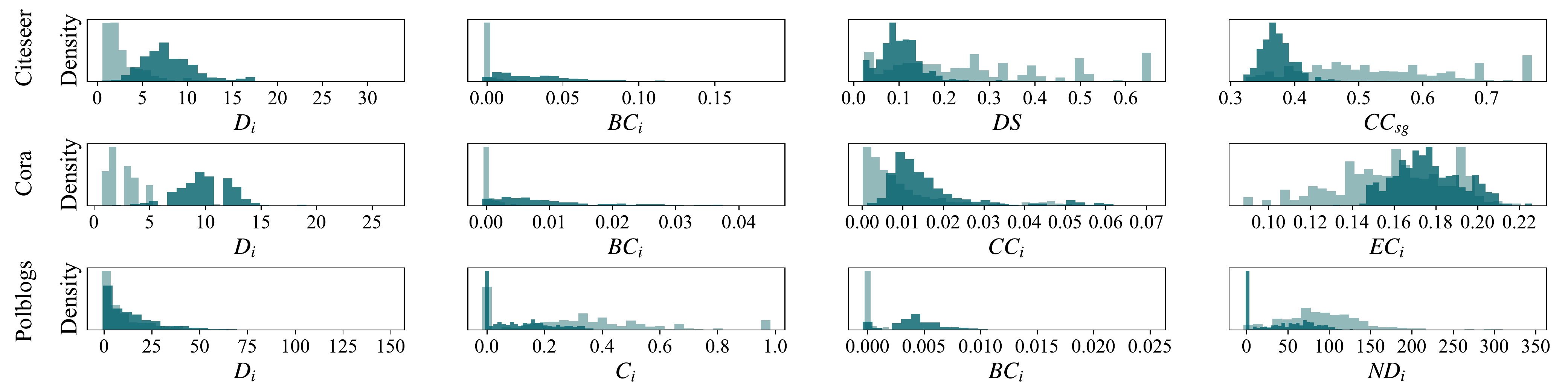}
\end{minipage}%
}%

\subfigure[GradArgmax]{
\begin{minipage}[t]{\linewidth}
\centering
\includegraphics[width=7in]{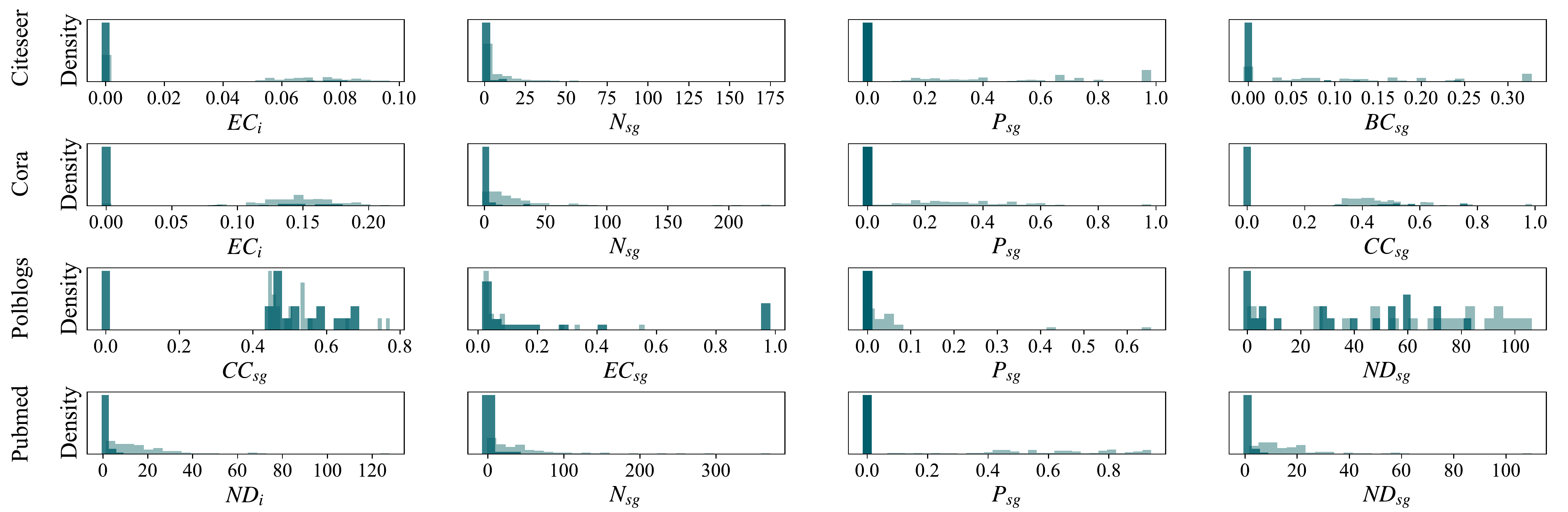}
\end{minipage}
}%

\subfigure{
\begin{minipage}[t]{\linewidth}
\centering
\includegraphics[width=2.5in]{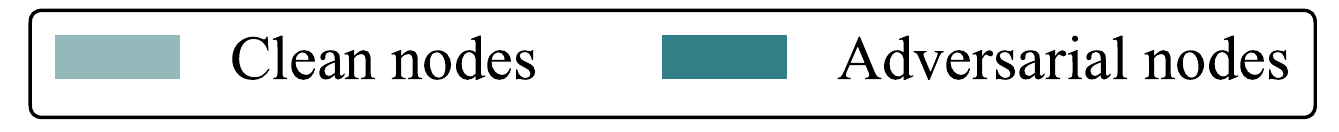}
\end{minipage}
}%

\centering
\caption{Changes of the top 4 most sensitive attributes when the graphs are disturbed by (a) Nettack, (b) Meta Attack, and (c) GradArgmax.}
\label{attr}
\end{figure*}

We considered the three adversarial attack methods in experiments, i.e., Nettack, Meta Attack, and GradArgmax, with their details presented in Section~\ref{sec:analysis}. The architectures and hyperparameters of the Nettack and Meta Attack are identical to those in \cite{zugner2018adversarial} and \cite{zugner2019adversarial}. As for GradArgmax, only two links are allowed to delete so as to ensure that the perturbations are imperceptible. Considering that attack patterns are more clearly reflected in the successful adversarial graphs, we only select adversarial graphs that can successfully fool the target model for analysis. 

\subsection{Attack Pattern Analysis}
For each adversarial attack on a network, we calculated the Gini Importance of all the 17 structural features on adversarial sample detection. The results are shown in Fig.~\ref{gini_impr}, where most graph attributes have quite low Gini Importance, indicating that all the three attack methods are only concealed to certain extent. However, there are still a few (3-5) graph attributes with relatively high Gini Importance for each attack method, which are relatively consistent across different networks, showing the distinct attack patterns. 

We summarize the top 4 graph attributes for each attack method on the four real-world networks, with results represented in TABLE~\ref{top3select}. We observe that Nettack has a great influence on $D_i$, $BC_i$, and $CC_{sg}$, Meta Attack mainly changes $D_i$ and $BC_i$, while GradArgmax mainly affects $P_{sg}$ and $N_{sg}$. This indicates that attributes affected by Nettack and Meta Attack are similar but completely different from those affected by GradArgmax, which is consistent with our theoretical results in Section~\ref{sec:analysis}: Nettack and Meta Attack have similar attack patterns since their attack goals are the same. Given each attack, we further visualize the distributions of the top 4 graph attributes for clean graphs and adversarial graphs, by Fig.~\ref{attr}. The histogram with darker color shows the distribution of attributes in the adversarial graph, and that with light color shows the distribution of attributes in the clean graph. Clearly, most of these distributions show significant horizontal shifting, showing that the concerned attributes indeed change much when the network structure is disturbed by the attack.  

\subsubsection{Pattern of Nettack}
The first attack pattern inferred theoretically is that Nettack tends to add links between the target node and the nodes with small degrees. The observation is as follows. The three centrality metrics $D_i$, $BC_i$, and $EC_i$ increase, indicating that the target node tends to get more neighbors to increase its centrality in the network; while both $ND_i$ and $CC_{sg}$ decrease, suggesting that the target node is more likely to be connected to the nodes with small degrees and lower closeness centrality. The second attack pattern is that Nettack is more likely to increase the number of common neighbors between the target node and others. We indeed find that the clustering coefficient $C_{i}$ increases in the experiments, implying that the number of common neighbors between the target node and its neighbors tends to increase. Note that, there is a significant difference between Polblogs and the other three networks. The top 4 graph attributes in Polblogs are mainly related to the neighbors of the target node while the top 4 attributes in other networks are mainly related to the node degree. We believe that such difference may be attributed to the higher graph density of Polblogs. For the graphs with lower density, such as Citeseer, Cora, and Pubmed, there are many small-degree nodes, which are easier to realize attacks by adding links between the target node and the others, i.e., the first attack pattern is preferred. While for the graphs with higher density, such as Polblogs, Nettack tends to increase the number of common neighbors between the target node and the others, since it is not easy to find suitable small-degree nodes in this situation. 

\subsubsection{Pattern of Meta Attack}
It is inferred theoretically that Meta Attack is quite similar to Nettack. Indeed, we find that the top 4 most sensitive graph attributes under Meta Attack and those under Nettack are highly consistent, and the variation trends of these attributes also about the same. This phenomenon demonstrates that these two attack methods have almost the same attack patterns, i.e., Meta Attack also tends to add links between the target node and the nodes with small degrees, and increase the number of common neighbors between the target node and the others. Note that, compared with Nettack, the variation of graph attributes caused by Meta Attack is more dramatic, due to the different attack strategies they adopt. Such result verifies again the inference performance analyzed in Section~\ref{sec:analysis}: Meta Attack uses gradient information, which helps it find the fastest direction to achieve the goal, leading to a larger variation of graph attributes.

\subsubsection{Pattern of GradArgmax}
Different from Nettack and Meta Attack, the top 4 most sensitive attributes under GradArgmax attack are graph-level attributes. All of these attributes have a downward trend, largely due to the link deletion operation in GradArgmax. Particularly, the attribute values almost equal to zero on Citeseer, Cora, and Pubmed, which means that the majority of adversarial nodes are isolated. This phenomenon indicates the first attack pattern: GradArgmax will get a high attack success rate when the target nodes have low degrees. This is also proven on the Polblogs. Since Polblogs has a relatively high density, i.e., there are few small-degree nodes, the attack success rate of GradArgmax on this network is even lower than 10\%. By comparison, there are a larger fraction of adversarial nodes that are not isolated (see the distributions of $CC_{sg}$, $EC_{sg}$, and $ND_{sg}$), but still, their $P_{sg}$ approaches zero. This indicates that the 2-hop subgraph of the adversarial node does not contain leaf nodes. This again shows that GradArgmax tends to delete the links of the leaf nodes around the target node.

\begin{table*}[t]
\renewcommand{\arraystretch}{1.3}
\caption{Performances of adversarial sample detection models. One is based on the top 4 most sensitive attributes and the other is based on all the 17 attributes. By comparison, they present similar performances in terms of ACC, AUC, and Precision.}
\label{comparison}
\centering
  \begin{tabular}{c|c c|c|c c|c|c c|c|c c|c}
    \toprule[2pt]
    Attack Method & \multicolumn{12}{c}{Nettack} \\
    \hline
    Dataset & \multicolumn{3}{c|}{Citeseer} & \multicolumn{3}{c|}{Cora} & \multicolumn{3}{c|}{Polblogs} & \multicolumn{3}{c}{Pubmed} \\
    \hline
    & Top 4 & All & $Gain(\%)$ & Top 4 & All & $Gain(\%)$ & Top 4 & All & $Gain(\%)$ & Top 4 & All & $Gain(\%)$ \\
    \hline
    ACC & 0.811 & 0.819 & 0.986 & 0.818 & 0.825 & 0.881 & 0.911 & 0.913 & 0.230 & 0.822 & 0.831 & 1.12 \\
    \hline
    AUC & 0.810 & 0.819 & 1.11 & 0.819 & 0.825 & 0.733 & 0.912 & 0.914 & 0.241 & 0.809 & 0.825 & 1.92 \\
    \hline
    Precision & 0.801 & 0.816 & 1.90 & 0.790 & 0.811 & 2.67 & 0.890 & 0.881 & -1.00 & 0.761 & 0.7972 & 4.76 \\
    \hline\hline
    Attack Method & \multicolumn{12}{c}{Meta Attack} \\
    \hline
    Dataset & \multicolumn{3}{c|}{Citeseer} & \multicolumn{3}{c|}{Cora} & \multicolumn{3}{c|}{Polblogs} & \multicolumn{3}{c}{Pubmed} \\
    \hline
    & Top 4 & All & $Gain(\%)$ & Top 4 & All & $Gain(\%)$ & Top 4 & All & $Gain(\%)$ & Top 4 & All & $Gain(\%)$ \\
    \hline
    ACC & 0.912 & 0.919 & 0.768 & 0.970 & 0.971 & 0.103 & 0.961 & 0.969 & 0.780 & - & - & - \\
    \hline
    AUC & 0.910 & 0.919 & 0.934 & 0.969 & 0.971 & 0.124 & 0.961 & 0.968 & 0.791 & - & - & - \\
    \hline
    Precision & 0.886 & 0.895 & 1.07 & 0.972 & 0.978 & 0.545 & 0.958 & 0.957 & -0.052 & - & - & - \\
    \hline\hline
    Attack Method & \multicolumn{12}{c}{GradArgmax} \\
    \hline
    Dataset & \multicolumn{3}{c|}{Citeseer} & \multicolumn{3}{c|}{Cora} & \multicolumn{3}{c|}{Polblogs} & \multicolumn{3}{c}{Pubmed} \\
    \hline
    & Top 4 & All & $Gain(\%)$ & Top 4 & All & $Gain(\%)$ & Top 4 & All & $Gain(\%)$ & Top 4 & All & $Gain(\%)$ \\
    \hline
    ACC & 0.958 & 0.974 & 1.67 & 0.978 & 0.984 & 0.624 & 0.982 & 0.984 & 0.193 & 0.978 & 0.975 & -0.286 \\
    \hline
    AUC & 0.958 & 0.975 & 1.70 & 0.975 & 0.985 & 1.03 & 0.981 & 0.984 & 0.289 & 0.980 & 0.973 & -0.694 \\
    \hline
    Precision & 0.984 & 0.989 & 0.427 & 0.964 & 0.972 & 0.924 & 0.965 & 0.969 & 0.363 & 0.983 & 0.992 & 0.997 \\
    \bottomrule[2pt]
  \end{tabular}
\end{table*}

In summary, the top $4$ attributes with the highest Gini importance can well explain the attack patterns theoretically inferred in Section~\ref{sec:analysis}, positively answering RQ1.

\subsection{Adversarial Sample Detection} \label{sect:RQ2}
Based on the above analysis, it is reasonable believe that the graph attributes can help detect adversarial samples and further recognize which attack method is used.

\subsubsection{Performance Metrics}
In order to compare the detection performances of different methods, we used the following metrics.

\begin{itemize}
  \item \textbf{Accuracy (ACC).} Accuracy is one of the most common evaluation metrics, which measures the detection performance with the percentage of correctly classified samples over all samples.
  \item \textbf{Area Under Curve (AUC).} AUC is the area under the Receiver Operating Characteristic (ROC) curve. It is equal to the probability that a classifier ranks a randomly chosen positive instance higher than a randomly chosen negative one. AUC measures the ranking of predictions rather than absolute values, which is independent of the used models.

  \item \textbf{Precision.} It is defined as 
  \begin{equation}
  Precision = \frac{TP}{TP + FP},\label{Prec}
  \end{equation}
  where $TP$ is the number of true positive samples and $FP$ is the number of false positive samples. Here, we take adversarial nodes as positive samples and clean nodes as negative samples. Then, adversarial nodes that can be successfully detected are true positive samples, and the clean nodes wrongly classified as adversarial ones are false positive samples.
\end{itemize}

To compare the performances of the model using all the 17 attributes, denoted by $M^{all}$, and that of the model using only the top 4 most sensitve attributes, denoted by $M^{top}$, we define the relative difference between the two as follows: 
\begin{equation}
    Gain = \frac{M^{all}-M^{top}}{M^{top}} \times 100\%.
    \label{gain}
\end{equation}

The results are presented in TABLE \ref{comparison}, where one can see that $Gain$ is lower than 1\% in most cases. Particularly, in 5 cases, they are even negative, which indicates that the left 13 attributes contribute little or even noise to the detection. These results show that the top 4 attributes contain enough information to distinguish adversarial samples from clean ones, positively answering RQ2. 

\begin{figure*}[ht]
\centering
\includegraphics[width=6.5in]{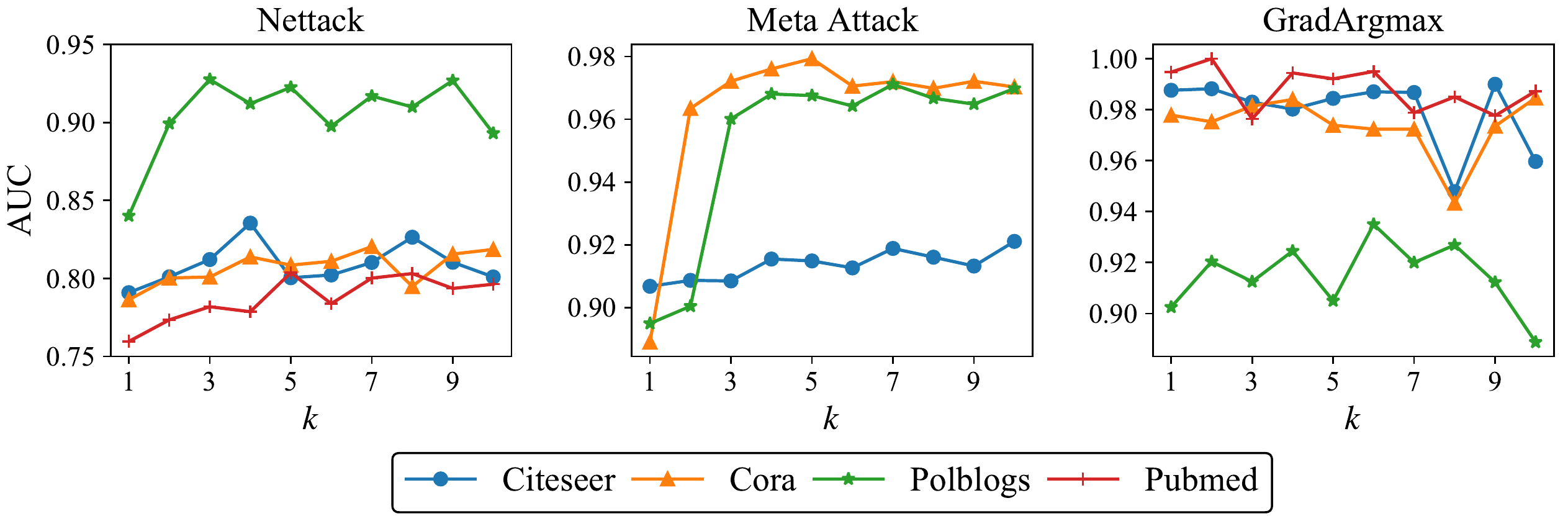}
\caption{AUC values for detecting adversarial samples when the top $k$ most sensitive attributes are adopted, with $k$ varied from 1 to 10.}
\label{k-sense}
\end{figure*}

In addition, we observe the impact of parameter $k$ on adversarial detection, as shown in Fig.~\ref{k-sense}. The results indicate that one can achieve promising detection performance using only one most sensitive attribute. The detection accuracy is above 0.75, around 0.9, and even as high as 0.98 in detecting adversarial samples generated by Nettack, Meta Attack, and GradArgmax, respectively. This suggests that all the three attack methods considered here are relatively \emph{simple}. By comparison, GradArgmax has the most clear attack pattern that can be captured by a single network feature. For Nettack and Meta Attack, the detection performance increases as $k$ increases from 1 to 4, and keeps steady as $k$ further increases, demonstrating that it is appropriate to choose the top 4 most sensitive attributes to establish the detection model. 

Note that, compared with the method in \cite{zhang2019comparing}, where the AUC values for detecting adversarial samples generated by Nettack and Meta Attack on Polblogs are around 0.86 and 0.70, respectively, our method achieves much better performance. Another point worth noticing is that those adversarial samples generated by gradient-based attack methods, i.e., Meta Attack and GradArgmax, are relatively easy to be detected in our framework.

\begin{table}[!t]
\renewcommand{\arraystretch}{1.3}
\caption{AUC values for recognizing adversarial attack methods}
\label{recog}
\centering
\begin{tabular}{c|c|c|c}
  \toprule[2pt]
  Datasets & Citeseer & Cora & Polblogs \\
  \hline
  MLP & $0.87\pm0.04$ & $0.84\pm0.05$ & $0.88\pm0.03$ \\
  \hline
  SVM (RBF) &$0.90\pm{0.03}$ &$\textbf{0.93}\pm{0.03}$ &$0.95\pm{0.01}$ \\
  \hline
  Random Forest & $\textbf{0.93}\pm{0.02}$ & $0.92\pm{0.02}$ & $\textbf{0.98}\pm{0.01}$ \\
  \bottomrule[2pt]
\end{tabular}
\end{table}

Since different adversarial attack methods have different attack patterns, and such patterns are similar across different networks, as shown in Fig.~\ref{gini_impr}, it is possible to further recognize the adversarial attack methods (as a multi-class task) based on the proposed network attributes. In this work, we leverage several common classifiers to address this problem. In addition to Random Forest, we also employ the Multilayer Perceptron (MLP) and Support Vector Machine (SVM) with Radial Basis Function (RBF) kernel, respectively. Since the computer capability is out of range when we perform Meta Attack on Pubmed, the experiments are conducted only on Citeseer, Cora, and Polblogs. 

The results are presented in TABLE \ref{recog}. By comparison, the Random Forest classifier achieves the best performance on two networks, with AUC values even higher than 0.92, while other classifiers also achieve comparable performance, indicating that our framework can well recognize adversarial attack methods based on their unique attack patterns.   

These results on adversarial samples detection and attack method recognition positively answer RQ3.

\section{Conclusion}\label{sec:conclusion}
In this paper, we investigate deeply several mechanisms to generate adversarial samples, and associate them with the variation of particular groups of network attributes. We find that Nettack and Meta Attack tend to modify the links between nodes with small degrees, and meanwhile increase the number of common neighbors between the target node and the others. GradArgmax, on the other hand, is more likely to attack those nodes with small degrees, leading to a lot of isolated adversarial nodes. Based on these findings, we further design machine learning algorithms based on a few network attributes for adversarial sample detection and attack method recognition, achieving well encouraging performances. 

As more and more adversarial attack methods emerge in recent years, in the future we will analyze more, and find the key network properties associated with them. Such research can provide interpretability for adversarial attacks in the area of network science and facilitate adversarial defense to protect graph data mining algorithms.


%

\ifCLASSOPTIONcompsoc
  \section*{Acknowledgments}
\else
  \section*{Acknowledgment}
\fi

The authors would like to thank all the members of the IVSN Research Group, Zhejiang University of Technology for the valuable discussions about the ideas and technical details presented in this paper.

\ifCLASSOPTIONcaptionsoff
  \newpage
\fi

\bibliographystyle{IEEEtran}
\bibliography{ref}

\end{document}